\documentclass[aps,prl,twocolumn,groupedaddress]{revtex4-1}

\usepackage{natbib}
\usepackage{graphicx}
\usepackage{dcolumn}
\usepackage{bm}
\usepackage{url}
\usepackage[caption = false]{subfig}
\usepackage{lipsum}
\begin{document}
\title{Angular Dependent Magnetization Dynamics with Mirror-symmetric Excitations in Artificial Quasicrystalline Nanomagnet Lattices}
\author{V.S.~Bhat$^{1,2}$ and D. Grundler$^{1,3}$} \email[]{vinayak.bhat@epfl.ch, dirk.grundler@epfl.ch} \affiliation{$^1$Laboratory of Nanoscale Magnetic Materials and Magnonics,  Institute of Materials (IMX), \'Ecole Polytechnique F\'ed\'erale de Lausanne (EPFL), 1015 Lausanne, Switzerland\\$^2$Lehrstuhl f\"{u}r Physik funktionaler Schichtsysteme, Physik Department E10, Technische Universit\"{a}t M\"{u}nchen, 85748 Garching, Germany\\ $^3$Institute of Microengineering (IMT), EPFL, 1015 Lausanne, Switzerland}

\vskip 0.25cm
\date{\today}

\begin{abstract}
We report angle-dependent spin-wave spectroscopy on aperiodic quasicrystalline magnetic lattices, i.e., Ammann, Penrose P2 and P3 lattices made of large arrays of interconnected Ni$_{80}$Fe$_{20}$ nanobars. Spin-wave spectra obtained in the nearly saturated state contain distinct sets of resonances with characteristic angular dependencies for applied in-plane magnetic fields. Micromagnetic simulations allow us to attribute detected resonances to mode profiles with specific mirror symmetries. Spectra in the reversal regime show systematic emergence and disappearance of spin wave modes indicating reprogrammable magnonic characteristics.
\end{abstract}

\maketitle
\begin{figure}
  	\includegraphics[width=0.49\textwidth]{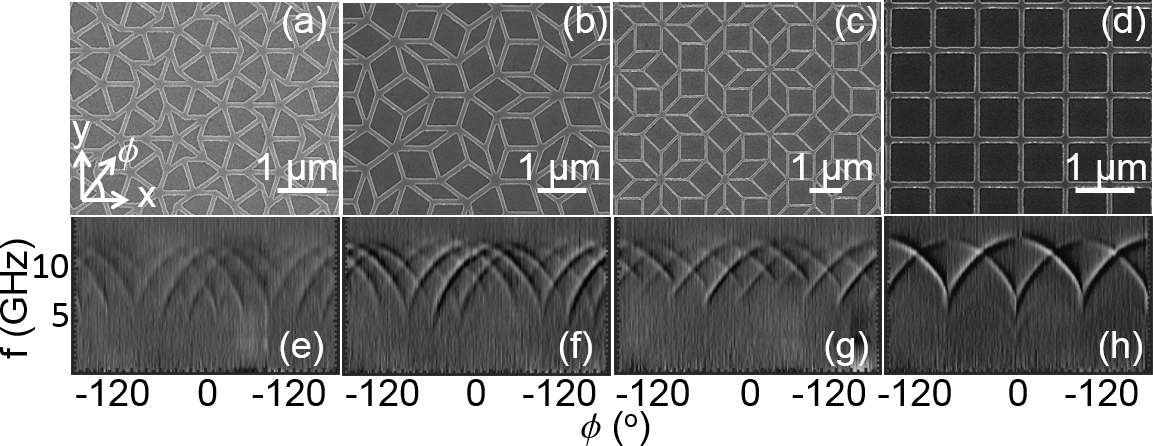}
  	\begin{flushleft}
  		\caption{Scanning electron microscopy images of inner sections of samples (a) P2T, (b) P3T, (c) AAT, and (d) SQT. Bright (dark) regions correspond to Py (GaAs). Gray-scale plots summarizing spin-wave absorption spectra measured at $\mu_{0}H$ = 100 mT as a function of angle $\phi$ for (e) P2T, (f) P3T, (g) AAT, and (h) SQT. To enhance the contrast we show difference spectra taken between neighboring datasets (derivative).}\label{Fig1}
  	\end{flushleft}
  \end{figure}
\begin{figure}
   	\includegraphics[width=0.445\textwidth]{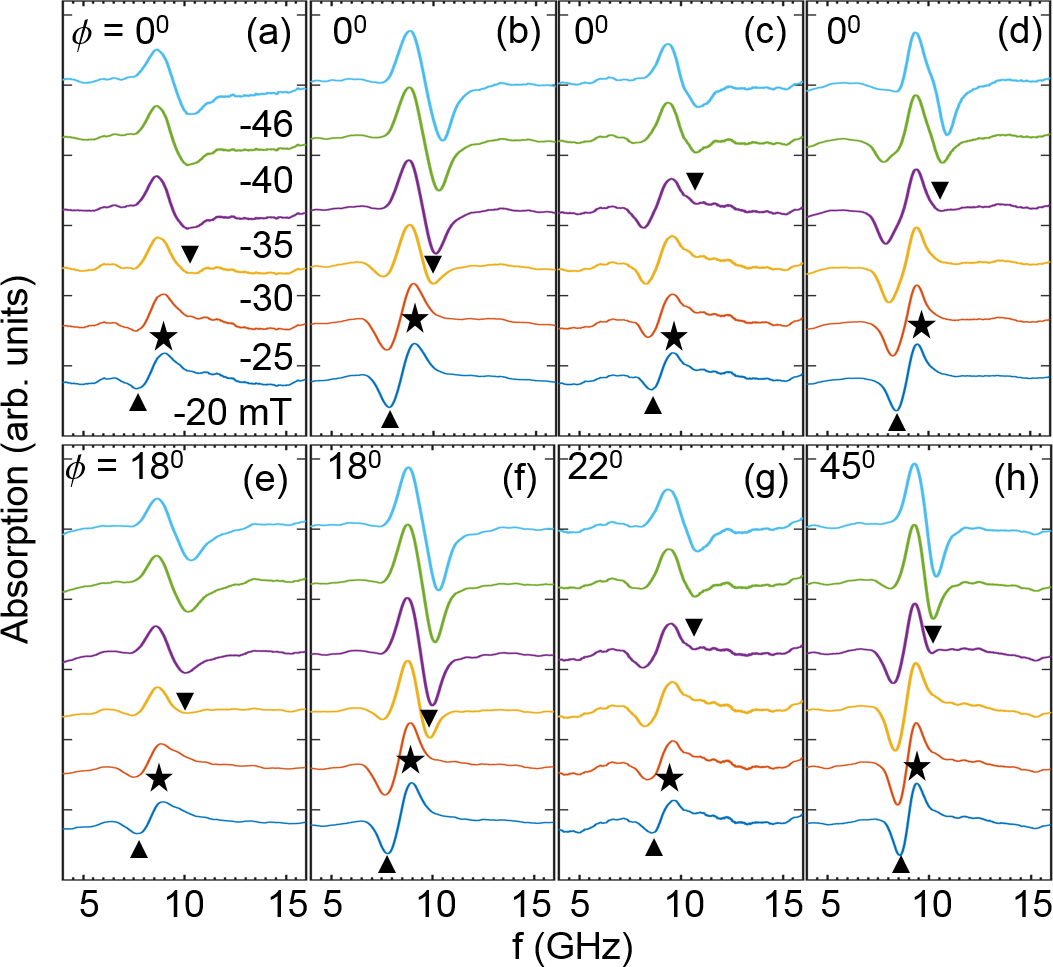}
   	\begin{flushleft}
   		\caption{Absorption spectra measured for different $\mathbf{H}$ at $\phi=0^{\circ}$ on (a) P2T, (b) P3T, (c) AAT, and (d) SQT. From each spectrum we subtracted the reference spectrum at $\mu_{0}H$ = 0 mT. The asterisks indicate the positive resonance peaks (maxima) belonging to the reference spectrum. Spectra taken on (e) P2T at $\phi=18^{\circ}$, (f) P3T at $\phi=18^{\circ}$, (g) AAT at $\phi=22^{\circ}$, and (h) SQT at $\phi=45^{\circ}$. Field values (labels) are given in mT and allocated to differently colored spectra in (a). They are valid for (a) to (h). The solid upward (and downward) triangles highlight resonances belonging to the main branches A before (and after the beginning of) the reversal. Notice the (re)appearance of the main mode with decreasing $H$ indicated by the downward triangles.}\label{Fig2}
   	\end{flushleft}
   \end{figure}
\begin{figure}
   	\includegraphics[width=0.49\textwidth]{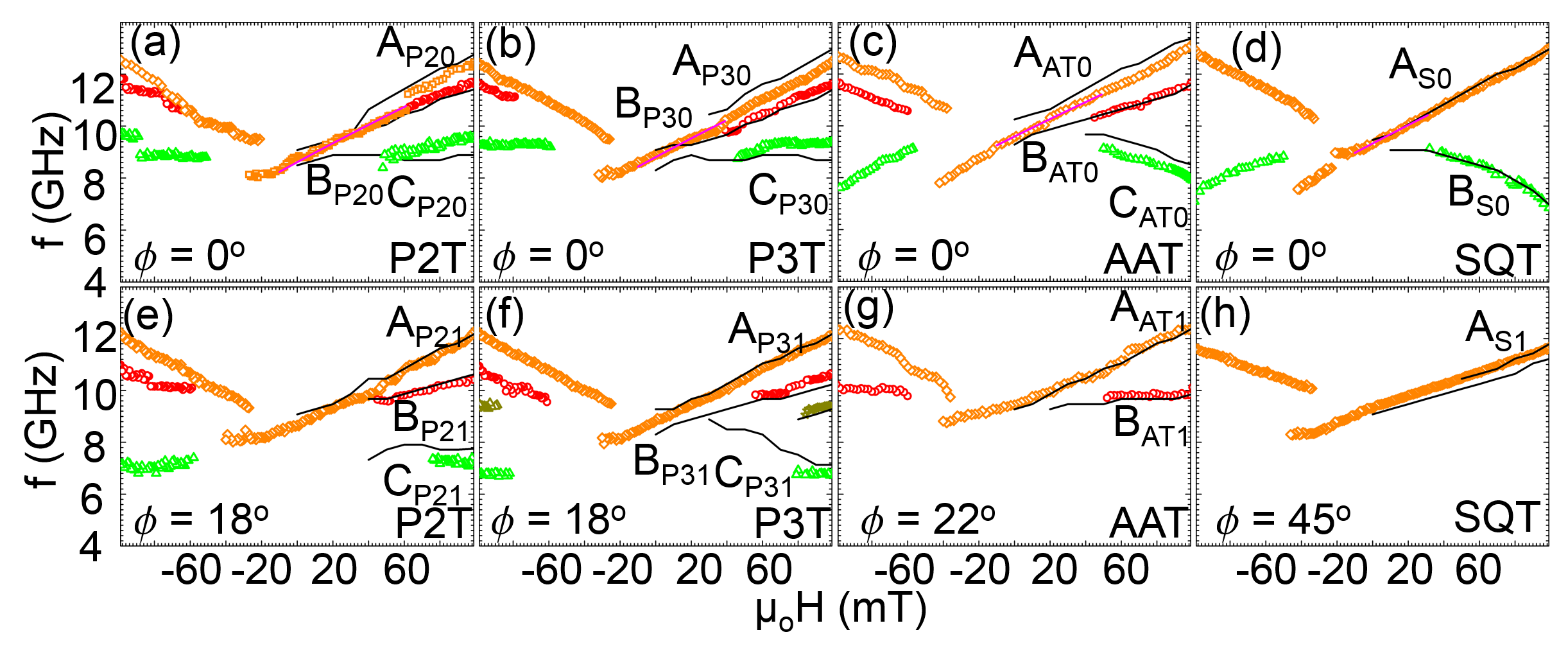}
   	\begin{flushleft}
   		\caption{Experimental resonance frequencies obtained for different applied field, $ H $, values for   (a) P2T at $\phi=0^{\circ}$, (b) P3T at $\phi=0^{\circ}$, (c) AAT at $\phi=0^{\circ}$, (d) SQT at $\phi=0^{\circ}$, (e) P2T at $\phi=18^{\circ}$, (f) P3T at $\phi=18^{\circ}$, (g) AAT at $\phi=22^{\circ}$, and (h) SQT at $\phi=45^{\circ}$. The applied field was varied from +100 mT to -100 mT. The solid black lines represent the simulated resonance frequency values. Magneta lines demonstrate Kittel formula fits to the branches representing spin wave modes at $\phi=0^{\circ}$. }   \label{Fig3}
   	\end{flushleft}
   \end{figure}

After thirty years of intense research on natural quasicrystals, that is, materials that exhibit long-range order but lack translational symmetry, one important question still challenges research: How does aperiodicity rule the physical properties that magnetic quasicrystals possess \cite{goldman2013family}? To address this question the material-by-design approach \cite{nisoli2013colloquium} was introduced recently in that two-dimensional ferromagnetic quasicrystalline tilings were patterned using nanolithography methods. Such materials were called artificial magnetic quasicrystals (AMQs). Using quasistatic techniques \cite{bhat2013controlled,bhat2014ferromagnetic, brajuskovic2016real,shi2018frustration} characteristic features of AMQs were reported such as knee anomalies in the magnetic hysteresis \cite{bhat2013controlled}, exotic low-energy configurations \cite{farmer2016direct, brajuskovic2016real, shi2018frustration}, and dendritic 2D avalanches of magnetically reversed segments \cite{brajuskovic2016real}. These features are not known from arrays where nanomagnets are arranged strictly periodically. They substantiate that nanobars arranged on aperiodic lattices interact. Preliminary studies on collective spin excitations in AMQs were presented in Refs. \cite{bhat2013controlled,bhat2014ferromagnetic} in which an in-plane magnetic field $\mathbf{H}$ was applied to saturate Penrose P2 and Ammann quasicrystalline tilings in one spatial direction. The role of magnetic disorder (or dendritic 2D avalanches) induced through partial reversal at intermediate fields $H$ has not yet been elucidated. In case of strictly periodic but frustrated lattices such as artificial spin ice magnetic configurations at intermediate fields showed characteristic spin-wave resonances \cite{bhat2016magnetization,bhat2017angular,jungfleisch2016dynamic,zhou2016large} that offered a novel approach to explore avalanche phenomena via microwave assisted switching \cite{bhat2016magnetization}. A similar study is lacking for an artificial magnetic quasicrystal though it forms an interesting candidate for magnonic crystals \cite{rychily2017spin} that might be reprogrammable \cite{krawczyk2014review,heyderman2013artificial}.\\ \indent In this Letter, we present a comprehensive experimental study and simulations on the collective spin excitations in three two-dimensional planar quasicrystals like aperiodic Penrose P2, P3 and Ammann tilings of different rotational symmetry [Fig. \ref{Fig1}(a) to (c) and Tab. \ref{tab:5/tc}]. We study them in magnetic fields $\mathbf{H}$ applied in different spatial directions. Comparing experimental spectra on quasicrystals with micromagnetic simulations and data obtained on a periodic lattice consisting of nominally identical nanomagnets [Fig. \ref{Fig1}(d)] we identify and categorize spin wave modes that occur in the quasicrystals. Our results open the pathway to reprogrammable magnonics with quasicrystals.
 \begin{table}[h]
	\centering
	\begin{tabular}{|p{1.3cm}|p{1.45cm}|p{1.45cm}|p{1.25cm}|p{1.1cm}|p{1.2cm}|}
		\hline
		Lattice & Lattice Symmetry & Magnetic Symmetry  & Main Branch &	$\mu_{0}H_{B}$ (mT) &$\mu_0N_xM_{\rm S}$ (mT) \\
		\hline
		Penrose P2 & 5-fold & 10-fold & $A_{P20}$ & 2.98 $\pm$ 0.35 & 24.4 \\
		\hline
		Penrose P3 & 5-fold & 10-fold & $A_{P30}$ & 8.2 $\pm$ 0.37 & 24.4\\
		\hline
		Ammann & 8-fold & 8-fold & $A_{AT0}$ &  8.39 $\pm$ 0.38 & 22.4 \\
		\hline
		Square & 4-fold & 4-fold & $A_{S0}$ & 22.7 $\pm$ 0.19 & 24.4\\
		\hline
	\end{tabular}
	\caption{\label{tab:5/tc} Bias fields evaluated from branches A using Eq. (\ref{eq:kittel}). Considering the symmetry-breaking quality of the applied in-plane magnetic field, the magnetic symmetries observed in Fig. \ref{Fig1} agree well with the expected characteristics.}
\end{table}\\
\indent Large (2.4 x 2.4 mm$^{2}$) lattices of nanomagnets were patterned on tilings representing Penrose P2 (P2T) \cite{bhat2013controlled}, Penrose P3 (P3T) \cite{gardner1997penrose,preshing2011}, Ammann (AAT) \cite{bhat2014ferromagnetic}, and a square (SQT) lattice (Fig. \ref{Fig1}) using nanofabrication techniques (see supplementary information for details). The length and thickness of a given Ni$_{80}$Fe$_{20}$ (Py) segment were 810 nm and 25 nm, respectively; the nominal width of a given nanobar for P2T, P3T, AAT, and SQT was 130, 130, 100, and 130 nm, respectively.  Assuming isolated nanobars and disregarding the vertices we estimated relevant demagnetization factors $N_{x} $ (along the long axis), $N_{y} $ (across the width), and $N_{z} $ (in out-of-plane direction) using Ref. \cite{aharoni1998demagnetizing}. Considering  $N_{x}+N_{y}+N_{z}=1$ we get values $N_{x}=0.02775$, $N_{y}=0.1841$, and $N_{z}=0.78815$ for the 810 nm long nanobars forming P2T, P3T, and SQT. For AAT, the values read: $N_{x}=0.02552$, $N_{y}=0.221$, and $N_{z}=0.75348 $. We note that the magnetic shape anisotropy fields $\mu_0H_{\rm ani}$ for the 810 nm long nanobars with widths of 130 nm and 100 nm amount to 137 mT and 172 mT, respectively \cite{o2000modern}. Nanobars of length 500 nm used in P2T [Fig. \ref{Fig1} (b)] exhibit $N_{x}$ = 0.044, $N_{y}$ = 0.179, and $N_{z}$ = 0.777. Room-temperature broadband spin-wave spectroscopy was performed in a flip-chip configuration \cite{bhat2016magnetization} (see supplementary information for details). Simulations were performed using the OOMMF code \cite{OOMMF1,bhat2016magnetization} on finite-size quasicrystals (see supplementary information for details). \\
\indent Eigenfrequencies measured at a fixed field value $\mu_{0}H=100$~mT for different angles $\phi$ are shown in Fig. \ref{Fig1} (e) to (h). The experimental data show prominent (main) branches for each of the four samples (black/white contrast). They exhibit ten-, ten-, eight-, and four-fold rotational symmetry, respectively, which deviates from the lattice symmetry in case of P2 and P3 (Tab. \ref{tab:5/tc}). Branches are slightly hysteretic consistent with $H<H_{\rm ani}$. Local frequency maxima in Fig. \ref{Fig1} (e) to (h) indicate that for the corresponding angle $\phi$ the magnetization vectors $\mathbf{M}$ of a subgroup of nanobars are parallel with both $\mathbf{H}$ and their easy axis. In this field orientation such nanobars exhibit the maximum internal field that enters the equation of motion and governs the precession frequency of spins \cite{gurevich1996magnetization}. Maximum frequencies are found to amount to about 12 to 13 GHz at 100 mT for all samples. Figure \ref{Fig2} shows spectra taken at fixed $\phi$ when $\mu_{0}H$ was decreased from +100 mT in a step-wise manner. For each lattice we depict spectra when field $\mathbf{H}$ was applied in two different directions.
In Fig. \ref{Fig2} (a) to (d) a pronounced single mode is seen in each lattice (local minima highlighted by upward triangle) when $\mu_{0}H$ is reduced to a small negative field of -20 mT (bottom-most spectra). The resonance frequency and amplitude of this mode decrease if $H$ is further diminished (from bottom to top). At a certain negative field value, a resonance (re)appears in each data set (downward triangle) whose signal strength and frequency steadily increase with more negative $H$. The emergence of such a high frequency mode indicates that nanobars which were initially aligned against the negative field direction have reversed their magnetization. At a field of -65 mT the high-frequency mode reaches a saturated  signal strength which might indicate that irreversible switching processes were completed. When we analyzed spectra obtained at the same field value $\mu_{0}H$ in successive field sweeps from +100 mT to -100 mT and back we observed reproducible resonance features when $\mu_{0}H$ resided in the regime of irreversible processes (Fig. S2 in the supplementary information). Reproducible spectra in disordered magnetic configurations are a prerequisite for reprogrammable magnonics. We attribute the reproducibility to nonstochastic switching \cite{bhat2014non} of identical nanobars enabled by a self-biasing effect in the quasicrystalline lattices due to different local environments. For photonic quasicrystals it was argued that an identical single defect produced different localized states depending on its specific placement and local dielectric environment \cite{Vardeny2013}. In the following we will argue that spin-wave resonance frequencies of nanobars monitor different magnetic environments in analogy to the defect in a photonic quasicrystal.\\ \indent In Fig. \ref{Fig3} we present the field-dependent resonance frequencies (symbols) which we extracted from large sets of spectra taken under conditions identical to Fig. \ref{Fig2}. Besides the main mode, we identified further resonances (branches) of small signal strength. We encode branches in three different colors with labels A, B and C.
For P2T shown in Fig. \ref{Fig3}(a) and \ref{Fig3}(e), we see three branches at +100 mT. In Fig. \ref{Fig3}(a) for $\phi$ = 0$^{\circ}$, we label them $A_{P20}$, $B_{P20}$, and $C_{P20}$. The branches $A_{P20}$ and $B_{P20}$ merge with each other for decreasing $\mu_{0}H$ at around 36 mT. The branch $C_{P20}$ exists down to 29 mT. At $H=0$ only one branch is resolved which dies out for negative fields. At about -23 mT (onset field), branch A reappears. This high-frequency branch is attributed to magnets with $\mathbf{M}$ parallel to both $\mathbf{H}$ and their easy axis direction. Branches B and C (re)emerge at -65 mT and -54 mT, respectively. We attribute the different onset fields to reversal fields of subgroups of nanobars that exhibit specific misalignment angls with $\mathbf{H}$. All onset field values are smaller compared to $\mu_0H_{\rm ani}$ indicating that reversal does not take place via coherent rotation \cite{Coey}. The observation of the three branches at $\mu_0H<-65~$mT is taken as an indication that the nanobars forming P2T have reversed to a large extent.\\ \indent In Fig. \ref{Fig3}(e) for $\phi$ = 18$^{\circ}$, we label the three distinct branches by $A_{P21}$, $B_{P21}$, and $C_{P21}$. The frequency separation between branches $A_{P21}$ and $B_{P21}$ is larger compared to $A_{P20}$ and $B_{P30}$ at 100 mT. Hence nanobars exhibit a larger variation in internal fields for this angle $\phi$. Branches $A_{P21}$ and $B_{P21}$ merge into a single branch at about 40 mT. This branch further decreases with decreasing $\mu_{0}H$. At -27 mT, -58 mT and -60 mT three branches successively appear that show a mirrored behavior compared to $A_{P21}$,  $B_{P21}$, and $C_{P21}$ at large positive $H$. Several branches are extracted for P3T [Fig. \ref{Fig3}(b) and (f)] and AAT [Fig. \ref{Fig3}(c) and (g)] as well. There exist however characteristic discrepancies: branches A to C change their slopes $df/dH$ from lattice to lattice. At the same time the onset field values are different. Comparing eigenfrequencies at around +40 mT branches A and B are found to split more and more in frequency from lattice to lattice (left to right in Fig. \ref{Fig3}).\\ 
\indent In Fig. \ref{Fig3} (d) and (h) we show the results obtained for the periodic square lattice SQT. Strikingly for $\phi=45^\circ$ in (h) only a single branch is resolved at 100 mT. Here all the nanobars experience the same misalignment with respect to the applied field $\mathbf{H}$ (45 deg). Correspondingly, the internal fields and eigenfrequencies are the same for all nanobars. In Fig. \ref{Fig3} (d) at $\phi=0^\circ$ two branches A and B of opposing slopes are identified for SQT. The positive (negative) slope $df/dH$ at $H>0$ is attributed to nanobars being collinear (orthogonal) to $\mathbf{H}$ \cite{gurevich1996magnetization}.\\
\indent In the following we present an analysis of spectra in that we make use of the demagnetization factors and model the most prominent branches in terms of uniform precession in a nanobar. By this means, we discuss reasons behind different slopes $df/dH$ of branches A focusing on $\phi = 0^\circ$. We assume that the local environment of nominally identical nanobars induces a (self-)bias magnetic field $H_{B}$. To estimate this field, we consider the Kittel equation for an individual magnetic ellipsoid in which we introduce $H_B$ as an additional magnetic field \cite{gurevich1996magnetization}
 {\begin{equation}\label{eq:kittel}
 \resizebox{0.5\textwidth} {!}{$f = \frac{\gamma}{2\pi}  \sqrt{\Big [ \mu_{0} (H+H_{B}) + \left( N_{z} - N_{x}  \right)\mu_{0} M  \Big ] \allowbreak \Big [ \mu_{0} (H+H_{B}) + \left( N_{y} - N_{x}  \right) \mu_{0} M  \Big ]   }.$}
 \end{equation}
Here, $H_{B}=0$ would represent an isolated nanobar without interacting neighbors. Its internal field is ruled by the demagnetization effect and its eigenfrequency is described by the unmodified Kittel equation. $H_{B}= N_\times M_s$ indicates that neighboring magnets compensate for the demagnetization field along a nanobar's long axis (i.e., they cancel magnetic charges) and induce a quasistatic internal field of zero similar to an infinitely long stripe. We fit Eq. (\ref{eq:kittel}) to branches A in Fig. \ref{Fig3}(a) to (d) (magenta lines behind the blue symbols for $H>0$). Using Eq. (\ref{eq:kittel}) we can quantitatively model the branches. Table \ref{tab:5/tc} summarizes the bias magnetic field $ \mu_{0}H_{B} $ evaluated for nanobars inside the four different lattices. The limiting case of $H_{B}= N_\times M_s$ is nearly fulfilled for SQT where nanobars collinear with $H$ form stripe-like chains. A considerably smaller value for the self-biasing field $H_{B}$ is extracted in case of P2T. Here, the local environment only weakly compensates for the demagnetization field. The eigenfrequency comes close to uniform precession of an isolated nanobar. Values $H_{B}$ are found to vary from quasicrystal to quasicrystal. Quasicrystals of the same rotational symmetry are found to exhibit clearly different bias fields. For the eight-fold symmetric Ammann lattice we find the largest value for a quasicrystal in Tab. \ref{tab:5/tc}. Our results suggest that local magnetic environments of nominally identical nanobars can be parametrized by $H_{B}$. The local bias field can explain nonstochastic switching and thereby allows for magnonic crystals with reprogrammable high-frequency responses.\\
\indent To understand in detail the microscopic origin of spin wave mode branches found in Figs. \ref{Fig2} and \ref{Fig3} in the saturated regime, we performed OOMMF simulations \cite{OOMMF1} in which we considered different in-plane angles $\phi$. Spectra simulated for $\mu_0H=100$~mT are depicted in Fig. \ref{illuMap}(a) for the four different samples. Here peaks indicate resonance frequencies . Simulations performed at many different field strengths and field orientations (see Fig. S3(a) for power spectra at different in-plane angles) provided us with resonance frequencies that we summarize as black lines in Fig. \ref{Fig3}. We find a good agreement for the number of branches between experiment and simulation.
\begin{figure}
	\includegraphics[width=0.5\textwidth]{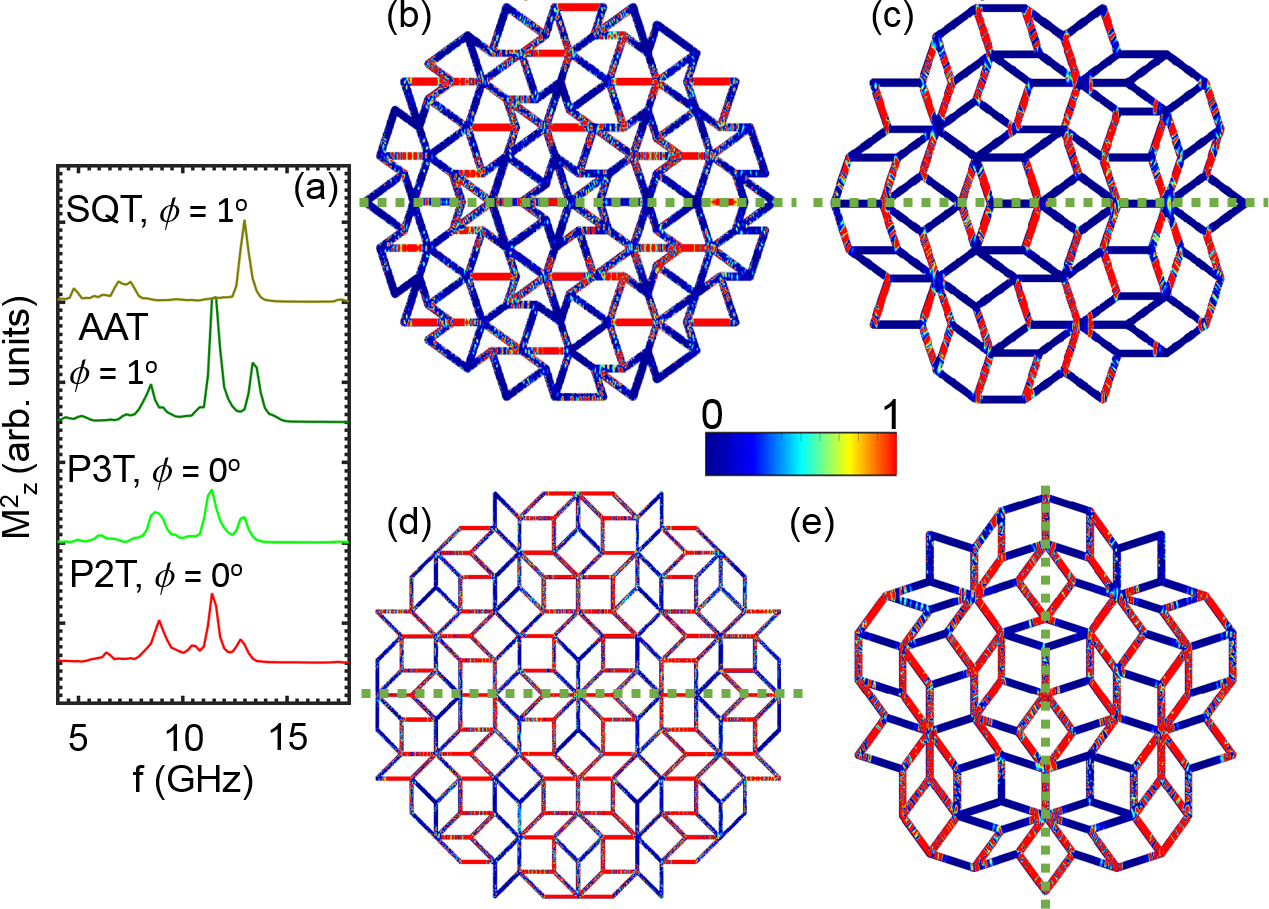}
	\begin{flushleft}
		\caption{  (a)  Simulated power spectra as a function of frequency for different lattices and in-plane angles at the external field value of $\mu_{0}H$ = 100 mT. The labels above each curve indicate the respective lattice and the in-plane angle, $\phi$.  Simulated local power map for the saturated state at $H=100$~mT and $\phi=0^{\circ}$ for (b) P2T  at $f$ = 12.79 GHz, (c) P3T  at $f$ = 10.25 GHz, (d) AAT  at $f$ = 13.57 GHz, and $\phi = 18^{\circ}$ for (e) P3T at $f$ = 8.69 GHz. The green dotted lines represent mirror axes. The magnetic field is along the horizontal direction.} \label{illuMap}
	\end{flushleft}
\end{figure}
Considering the consistency we show spatial distributions of spin-precessional amplitudes in Fig. \ref{illuMap} (b) to (d) which we attribute to the modes of branches A at $\phi=0^\circ$. Here, the magnetization of nanobars collinear with the $x$-direction (field direction) is found to precess pronouncedly (red). Their precession is largely uniform supporting the modelling based on the Kittel formula (Eq. \ref{eq:kittel}). We clearly observe that the modes in Fig. \ref{illuMap} (b) to (d) are mirror-symmetric with respect to axes shown with green dotted lines. The axes are parallel to $\mathbf{H}$. Surprisingly, when the field is applied along an off-symmetry axis ($\phi = 18^{\circ}$ for P3T), we again observe the existence of a mirror axis [Fig. \ref{illuMap} (e)]. However, now the axis is perpendicular to $\mathbf{H}$. Further power maps on the quasicrystals suggest that branches B and C at large $\mu_{0}H$ can be interpreted as follows: for P2T, $B_{P20}$, and $C_{P20}$ arise from nanobars with angles $\pm$ 36$^{\circ}$, and $\pm$72$^{\circ}$, respectively, with respect to $\mathbf{H}$ consistent with the observed ten-fold rotational symmetry. $B_{P21}$ of P2T correlates to spin precession in nanobars oriented at $\pm 54^{\circ}$. Similar allocations hold true for P3T. For AAT and $\phi$ = 0$^{\circ}$, branches  $A_{AA0}$, $B_{AA0}$, and $C_{AA0}$  seem to belong to nanobars at $\phi$ = 0$^{\circ}$, $\pm$ 45$^{\circ}$,  and $\pm$90$^{\circ}$, respectively. For $\phi$ = 22$^{\circ}$ two spin wave modes $A_{AA1}$ and $B_{AA1}$ arise from nanobars making an angle of $\phi$ = $\pm$ $22^{\circ}$, and  $\pm$68$^{\circ}$, respectively, with respect to $\mathbf{H}$. Overall prominent branches observed at the distinct angles $\phi$ considered in Figs. \ref{Fig2} and \ref{Fig3} reflect spin precession in subgroups of nominally identical nanobars which due to the long-range order are oriented in mirror symmetry under specific angles relative to $\mathbf{H}$. Their exact eigenfrequency is governed by the local environment of the subgroups. In Fig. \ref{illuMap} (d) the prominently excited nanobars form band-like patterns extending in a direction perpendicular to $\mathbf{H}$ applied along $x$-direction. These bands do not exhibit translation symmetry along $x$-direction due to the underlying quasicrystalline lattice. Further studies are required to understand how mode profiles vary when magnetic disorder and topological defects \cite{gliga2013spectral} are present. We note that in the regime of irreversible switching we found reproducible spectra for accordingly disordered quasicrystals (Fig. S2 in the supplementary information). Our results presented here provide the basis for studies addressing quasicrystals as exotic artificial spin ice structures incorporating topological defects \cite{farmer2016direct}\\
\indent To summarize, we  studied artificial quasicrystalline ferromagnets based on Penrose P2, P3, and Ammann tilings. We observed systematic variations and reproducible series of field-dependent resonance frequencies across the hysteresis loops. The detailed comparison between experiment and simulations in the saturated regime indicates that the shape anisotropy of the individual nanobars played a dominant role for the value of the resonance frequency when considering the bias-field effect of the different long-range-ordered local environments. Penrose P2 and P3 tilings are found to exhibit striking similarities concerning spin wave modes residing in nanobars with orientations that are mirrored with respect to the applied field. Resonances in the Ammann tiling could be explained along a similar line assuming correspondingly different angles. Our experiments and findings pave the way for studies on reprogrammable magnonics based on quasicrystals and the spin dynamics of topological defects in exotic artificial spin ice.
 \begin{acknowledgments}
	The research was supported by the Swiss National Science Foundation via Grant No. 163016. V. S. Bhat thanks B. Farmer and I. ~Stasinopoulos for their help in editing the python code for Penrose P3 tiling in Ref. \cite{preshing2011} to generate dxf files, and for the technical help in fabrication of CPW, respectively.
\end{acknowledgments}

\end{document}